\documentclass[sigconf]{acmart}

\AtBeginDocument{%
  }

\renewcommand\footnotetextcopyrightpermission[1]{}

\usepackage{fancyhdr}
\usepackage{array}
\usepackage{multirow}
\usepackage{graphicx}
\usepackage{textcomp}
\usepackage{algorithm, algpseudocode}
\usepackage{xcolor}
\usepackage{soul}
\usepackage{algorithm}
\usepackage{algpseudocode}

\begin{document}

\fancyhead{}
\title{The Quantum Imitation Game: Reverse Engineering of Quantum Machine Learning Models}

\author{Archisman Ghosh}
\email{apg6127@psu.edu}
\orcid{0000-0002-0264-6687}

\affiliation{%
  \institution{Pennsylvania State University}
  \city{State College}
  \state{PA}
  \country{USA}
}

\author{Swaroop Ghosh}
\email{szg212@psu.edu}
\affiliation{%
  \institution{Pennsylvania State University}
  \city{State College}
  \state{PA}
  \country{USA}
}








\begin{abstract}
Quantum Machine Learning (QML) is an amalgamation of quantum computing paradigms with machine learning models, providing significant prospects for solving complex problems. However, with the expansion of numerous third-party vendors in the Noisy Intermediate-Scale Quantum (NISQ) era of quantum computing, the security of QML models is of prime importance, particularly against reverse engineering, which could expose sensitive parameters and proprietary algorithms embedded within the models. We assume the untrusted third-party quantum cloud provider is an adversary having white-box access to the transpiled version of the user-designed trained QML model during inference. Although the adversary can steal and use the model without any modification, reverse engineering (RE) to extract the pre-transpiled copy of the QML circuit will enable re-transpilation and usage of the model for various hardware with completely different native gate sets and even different qubit technology. Such flexibility may not be obtained from the transpiled version of the circuit which is tied to a particular hardware and qubit technology. The information about the parameters (e.g., number of parameters, their placements, and optimized values) can allow further training of the QML model if the adversary plans to alter the QML model to tamper with the watermark and/or embed their own watermark or refine the model for other purposes. In this first effort to investigate the RE of QML circuits, we examine quantum classifiers by comparing the training accuracy of original and reverse-engineered models across various sizes (i.e., number of qubits and number of parametric layers) of Quantum Neural Networks (QNNs). We note that multi-qubit classifiers can be reverse-engineered under specific conditions with a mean error of order $10^{-2}$ in a reasonable time. We also propose adding dummy rotation gates in the QML model with fixed parameters to increase the RE overhead for defense. For instance, an addition of 2 dummy qubits and 2 layers increases the overhead by $\sim1.76$ times for a classifier with 2 qubits and 3 layers with a performance overhead of less than 9\%. We note that RE is a very powerful attack model which warrants further efforts on defenses.
\end{abstract}





\maketitle

\section{Introduction}
\label{intro}
Quantum Machine Learning (QML) merges the cutting-edge capabilities of quantum computing with sophisticated machine learning techniques, offering the potential to solve complex problems intractable for classical computers \cite{schuld2015introduction}. QML circuits involve quantum properties like entanglement and superposition to explore the Hilbert space more effectively and thus are able to process and analyze vast amounts of data with enhanced speed and efficiency. However, with the advancement in QML design and the increase in the complexity of the models, there is an increased demand for quantum hardware. To cater to this increasing demand, quantum hardware providers have taken the initiative to provide QML hardware as a service to aid the design and utilization of advanced QML models. In the noisy intermediate-scale quantum (NISQ) era of quantum computing \cite{Preskill_2018}, the number of third-party cloud-based quantum hardware providers will only increase thus reducing the cost of using quantum hardware. One pressing issue is the potential incentive of some rogue adversary or an untrusted third-party cloud provider to steal trained QML circuit designs, posing significant threats to the privacy and integrity of these models \cite{suryansh_acm}.

\subsection{Why QML Models are at Risk}
QMLs face significant security risks due to the following reasons: 

\textbf{High training cost:} Quantum computers are expensive e.g., \$1.6 per second for IBM’s superconducting qubits and \$0.01 per shot for IonQ’s Trapped Ion (TI) qubits. This is at least $10^5 \times$ costlier than classical resources which is priced $\sim\$ 2.1 \times 10^{-6}$ per second. QML models require hundreds of training epochs each with thousands of quantum circuit executions (depending on the size of the training dataset). Each circuit is executed for thousands of trials to get expectation values. This makes the trained and even partially trained QML model very expensive. Compared to current state-of-the-art ML models e.g., Gemini that take millions to billions of dollars for training, QML models at scale may cost many orders of magnitude higher making them extremely valuable. 

\textbf{High training time:} Current state-of-the-art ML models e.g., ChatGPT3 took $\sim 1$ month for training using thousands of dedicated GPUs. Quantum resources are scarce whereas their demand is extensive. As a result, both hardware, as well as simulators (whose computation time scales exponentially with qubit size) hosted in the cloud, incur long wait queues. This is true for even dedicated access to quantum computers such as membership of a Quantum Hub with a small set of users. Therefore, training a large QML model might take a significant amount of time (e.g., months to years). 

\textbf{Hosting of QMLs on the quantum cloud:} Since QML providers may not possess their own quantum hardware, they may rely on a third-party quantum cloud for hosting the model. This will lead to the rise of QMLaaS \cite{10.1145/3649476.3658806} providing access to clients only through input-output queries via external APIs. The quantum cloud provider may have white-box access to the expensive model and training data. 

\textbf{Miscellaneous IPs:} The untrained QML IPs include model architecture (i.e., entanglement, number of parameters, number of layers, measurement basis) and training data embedded in state preparation circuit. The trained QML IPs include optimized parameters and input data embedded in the state preparation circuit during inference. 

\subsection{Attack Model and Motivation}
During inference operation, the input data is first appended as a state preparation circuit within the trained QML model. Next, the model is transpiled for target quantum hardware where logical qubits are mapped to physical qubits, $SWAP$ gates are added to meet the hardware coupling constraints and complex gates, and the trained parametric rotation gates are decomposed into native gates. Finally, the transpiled QML circuit is sent to the quantum cloud for execution. Access to the white-box architecture of the trained QML circuit will allow the untrusted cloud providers to potentially steal and use it. For example, the adversary can strip off the state preparation circuit to extract the trained portion of the QNN and attach their own input data for inference on the same target hardware. They can also sell the trained portion of the QNN. For such attacks, knowing the original QML model and optimized values of the parameters is not important.  

Nevertheless, knowing the original entanglement and optimized parameters can provide adversaries with several additional advantages such as: (i) enabling the transpilation and use of the model on different hardware platforms with varying native gate sets and qubit technologies. This flexibility might not be achievable with the hardware-specific transpiled version; (ii) revealing the entanglement architecture of the QNN, which can be sold separately or utilized to train a clone model with different datasets; (iii) providing detailed information about the parameters, including their optimized values and placements, facilitating further training or tampering with the QML model, such as watermark alteration or embedding. This threat is comparable to the usage of reverse engineering and decompilation techniques that are used by adversaries in classical hardware system design to bypass watermarking measures and counterfeiting architecture \cite{rehw}.

\subsection{RE and Associated Challenges}
Reverse engineering is the idea of analyzing a model and trying to recreate its design preserving its architectural nuances \cite{re1}. From the quantum perspective, it involves reconstructing the original hardware-agnostic quantum circuit from its optimized and hardware-specific transpiled form (Fig. \ref{fig:flow}). 
The reverse engineering of QML models presents unique challenges compared to classical machine learning (ML) models, particularly due to differences in model representation, transpilation, and hardware dependency. Classical ML models are represented as mathematical functions or neural networks, while QML models are depicted as quantum circuits with quantum gates as parameters. The transpilation procedure is necessary to convert the QML circuit design to fit the native gate set of the training and inferencing hardware. 
While finding the entanglement architecture from the transpiled circuit is relatively easy as the only task is to reverse logical to physical mapping while identifying and accounting for the $SWAP$ gates, we note that recovering the original parameters from decomposed and optimized single-qubit gates is non-trivial. The reasons are multi-fold, (i) transpilation converts all single-qubit gates to basis gates ($RZ(\theta)$ in the case of IBM machines). This makes identification of the original rotation gate type difficult, (ii) transpilation of parameterized rotation gates results in a sequence of single-qubit gates which gets optimized with rotation gates resulting from other parameterized or non-parameterized rotation gates from the QML model, (iii) the transpilation process often adds global phase during optimization to maintain a correct relative phase between the states of qubits which obfuscates the original parameters, (iv) increasing the optimization level selected during transpilation enforces stricter optimization rules adding another level of obscurity. These are covered in Section \ref{bg}. 


\emph{To the best of our knowledge, this is the first attempt to reverse engineer a quantum machine learning circuit.} Note, that our primary contribution is the extraction of trained parameters. The extraction of entanglement architecture and training data is beyond the scope of the paper. The major contributions are as follows:
\begin{figure}
    \centering
    \includegraphics[width=0.7\linewidth]{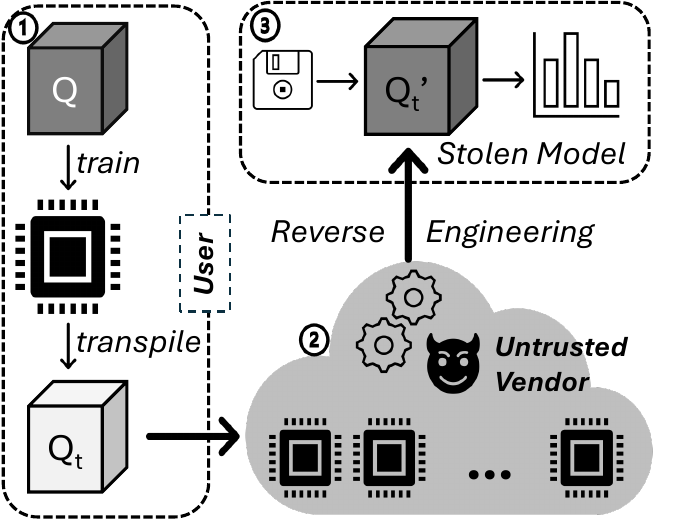}
    \caption{The flow diagram describes reverse engineering of QML parameters by untrusted third-party vendors acting as adversaries. (1) shows the user training and transpiling a QML model $Q$ using non-proprietary quantum hardware and sending the transpiled version of the trained model $Q_t$ to the untrusted vendor for inferencing. (2) and (3) describe the attack model involving the procedure of reverse engineering performed by the untrusted vendor to extract the parameters and steal the IP of the user-designed model.}
    \label{fig:flow}
    \vspace{-10pt}
\end{figure}

\begin{enumerate}
    \item We present a methodology for reverse engineering transpiled QML circuits.
    \item We propose a procedure for extracting the original parameters from transpiled QML circuits which can be used by the adversary to obtain a duplicate of the model.
    \item We demonstrate the efficacy of the proposed idea by reverse engineering multi-qubit classifiers.
    \item We perform an overhead analysis and discuss potential countermeasures.
\end{enumerate}

\subsection{Paper Structure}
Section II provides a background on quantum computing and the compilation of quantum circuits. Section III presents the threat model. Section IV presents the proposed reverse engineering procedure and Section V comprises a detailed study of the experiments. Section VI develops countermeasures and Section VII concludes the paper.
\section{Background}
\label{bg}

\subsection{Quantum Computing}
In quantum computing, the fundamental unit of computation is the quantum bit, or qubit. Unlike classical bits, which can be in one of two states (0 or 1), a qubit can exist in a superposition state, which is a linear combination of both 0 and 1. In Dirac notation, the state of a qubit is represented as $|\psi\rangle=\alpha|0\rangle+\beta|1\rangle$, where $\alpha$ and $\beta$ are complex coefficients that satisfy the normalization condition $|\alpha|^2+|\beta|^2=1$. Here, $|0\rangle = [1 \ 0]^T$ and $|1\rangle = [0 \ 1]^T$ are the computational basis vectors. Therefore, $n$ qubits can be used to represent a space of $n$-qubit states with $2^n$ basis states, ranging from $|0...0\rangle$ to $|1...1\rangle$, and an $n$-qubit state $|\psi_n\rangle$ can be represented as $|\psi_n\rangle = \sum_{i=0}^{2^n-1} a_i|i\rangle$; where $\sum_{i=0}^{2^n-1} |a_i|^2=1$.

 Quantum algorithms exploit superposition, entanglement, and quantum interference in qubits, to perform computations that can be exponentially faster than their classical counterparts for certain tasks.
\begin{figure}
    \centering
    \includegraphics[width=1\linewidth]{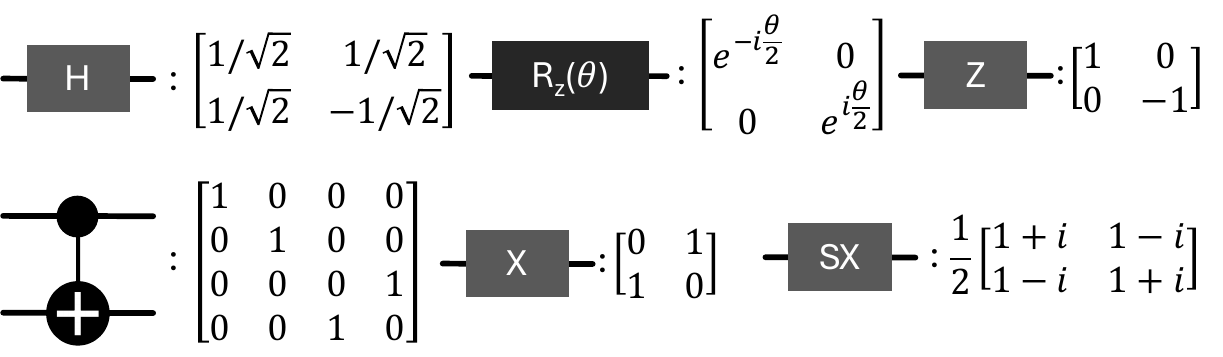}
    \caption{Matrix representation of basic quantum gates, Hadamard, Rotation-Z, Pauli-Z, $CNOT$, Pauli-X, and SX (from top left to right). An $n$-qubit gate is represented by a $2^n\times2^n$ matrix.  } 
    \label{fig:gates}
    \vspace{-10pt}
\end{figure}
Quantum logic gates that are analogous to classical logic gates are used to exploit the quantum properties by allowing the interaction between qubits. They are the building blocks of quantum circuits. Common quantum logic gates (Fig. \ref{fig:gates}) include the Pauli-X (NOT) gate, which flips the state of a qubit, the CNOT gate, which entangles pairs of qubits, Pauli-Z gate, which flips the phase of the qubit, and rotation gates which operate using parameter values and allow more precise control over the quantum states of the qubits. A quantum circuit is constructed by applying one or more gates in a sequence.

\subsection{Compilation of Quantum Circuits}

\begin{figure}
    \centering
    \includegraphics[width=1\linewidth]{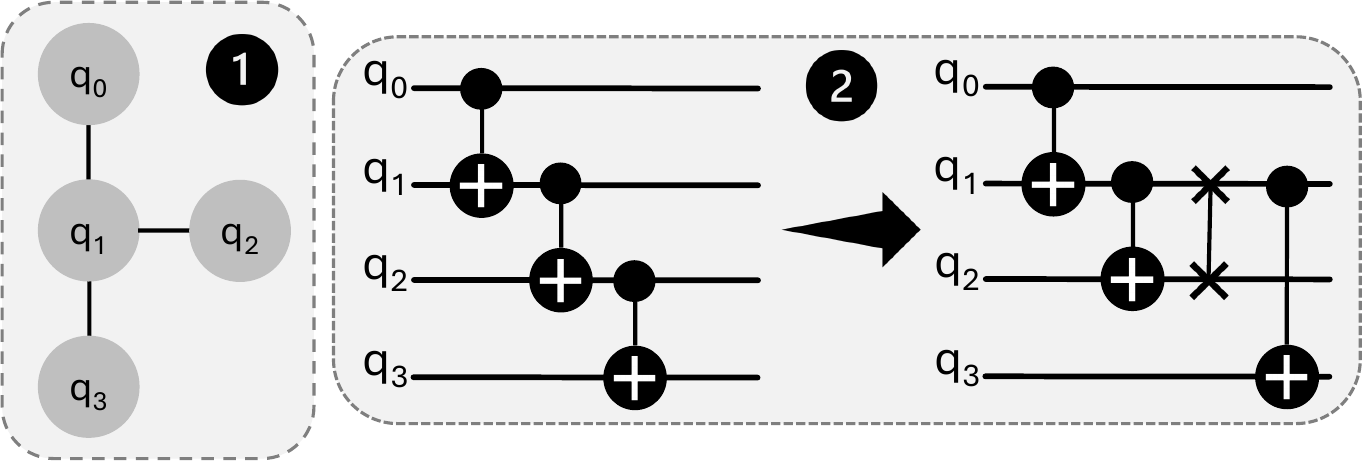}
    \caption{A diagrammatic representation of the $SWAP$ operation during transpilation of a quantum circuit. (1) represents the T-shaped coupling map of the quantum hardware where the circuit is transpiled and sent for execution. (2) shows the transpilation procedure where a $SWAP$ gate is inserted between $q_1$ and $q_2$ to accommodate the physical layout of the qubits on the quantum hardware.} 
    \label{fig:coup}
    \vspace{-10pt}
\end{figure}

Compilation of quantum circuits involves several steps to translate the high-level quantum programs to a form compatible with the particular constraints of quantum hardware. In IBM terminology, this is referred to as \textit{transpilation} \cite{trans}. \textbf{Gate Translation:}
Quantum programs are typically written using high-level gates, which are abstract representations of quantum operations. However, current quantum computers only support a limited set of native instructions known as basis gates. For example, IBM quantum machines support the following basis gates: \texttt{[id, x, sx, cnot, rz]}. Therefore, any high-level instructions in a quantum program must be translated into these native instructions to be executable on the hardware. This translation step is essential for aligning the abstract quantum algorithm with the practical limitations of the quantum hardware. \textbf{Coupling Map Constraints:}
In addition to the instruction set alignment, quantum hardware architectures face another significant challenge known as the coupling map constraint. This constraint arises from the physical layout of the qubits on the hardware. For instance, Fig. \ref{fig:coup} (1) demonstrates a T-shaped coupling map of quantum hardware of 4 qubits, where the nodes represent physical qubits. An edge between two nodes indicates that a two-qubit operation (such as a $CNOT$ gate) between those physical qubits is directly allowed. Suppose we have a sample three-qubit quantum program that we want to run on this quantum hardware. To execute the program, each logical qubit in the program must be mapped to a separate physical qubit on the hardware. However, the $CNOT$ gate between qubits q2 and q3 cannot be directly executed with this mapping because there is no edge in the coupling map between q2 and q3, as shown in Fig. \ref{fig:coup}. This is an example of a coupling constraint. To resolve this constraint, qubits must be routed using the SWAP operation so that logical qubits involved in two-qubit operations become nearest neighbors. A SWAP gate between q1 and q2 swaps the state of the physical qubits q1 and q2 allowing the $CNOT$ gate to be applied between q2 and q3 via q1. During the transpilation procedure, the $SWAP$ gate in Fig. \ref{fig:coup}(2), gets converted into a sequence of three $CNOT$ gates with single-qubit rotation gates in between. These rotation gates are not part of the original QML model (see Fig. \ref{fig:LUT}(2))

\subsection{Quantum Neural Networks}
\begin{figure}
    \centering
    \includegraphics[width=1\linewidth]{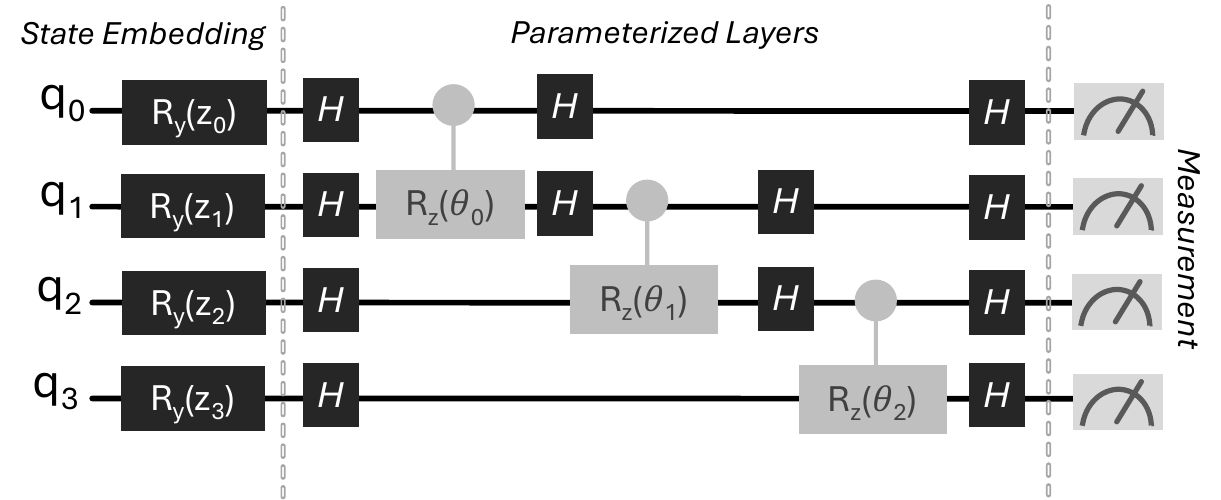}
    \caption{A circuit representation of a PQC. In state embedding, the $RY(z_i)$ gates are used for basis encoding to map the data to the computational basis states. The parameterized layers comprise a cascade of $CRZ(\theta_i)$ gates that provide the entanglement as well as a finer-grained search into the Hilbert Space. Measurement operators that follow, measure the outcome of individual qubits to derive an output. } 
    \label{fig:pqc}
    \vspace{-10pt}
\end{figure}

Quantum Neural Networks (QNNs) represent the intersection of quantum computing and machine learning \cite{Schuld2014}. Quantum circuits when designed in such a way that they embed classical data as states of qubits, can perform tasks similar to classical neural networks like regression and classification. \textbf{Quantum Data Encoding:}
This is the initial step in designing a QNN where quantum data encoding embeds the classical data in the Hilbert space through the quantum states of the qubits. Methods include amplitude encoding, which normalizes and encodes data into the amplitudes of qubits, angle encoding, which converts data into rotation angles applied to qubits, and basis encoding, which maps binary data directly to computational basis states. \textbf{Parameterized Quantum Circuits (PQCs): }
The core of a QNN is the Parameterized Quantum Circuit (PQC) (Fig \ref{fig:pqc}), consisting of adjustable quantum gates. PQCs include quantum rotation gates like $RX(\theta), RY(\theta)$, and $RZ(\theta)$ with tunable parameters. The circuit architecture defines qubit interactions, and entanglement between qubits enhances computational power. PQCs allow QNNs to perform complex transformations like convolution, akin to layers in classical neural networks. \textbf{Measurement: }
Measurement extracts classical information from quantum states after computation. Quantum states collapse upon measurement, revealing the final qubit state. The probability of each basis state is measured to derive outputs. Measurement results are processed classically, aggregating outcomes or applying post-processing techniques. The QNN workflow starts with preprocessing and encoding classical data into quantum states. The encoded data enters a PQC with initial gate settings. After processing, measurements extract classical information that is used to compute the loss function. Classical optimization algorithms are implemented alongside the QNNs to iteratively adjust the PQC parameters, minimizing the loss function until convergence which enables it to learn and improve performance.
\subsection{Related Work}

Reverse engineering attacks on convolutional neural networks (CNNs) running on hardware accelerators have been explored before \cite{revcnn}. It is shown in the paper that side-channel attacks on memory can help adversaries infer network structure and even the weights of the CNN in spite of data encryption. Dynamic zero-pruning in CNN accelerators can leak weight values as well which can be protected by hiding off-chip memory access patterns. There have been attempts to perform fault attacks to reverse engineer neural networks \cite{Breier_2022}. On introducing controlled faults in the neural network hardware, output changes can be observed to deduce the architecture of the model, thus highlighting the need to engineer fault-tolerant neural networks. 
Black-box neural networks in the classical domains can also be attacked by querying it and observing the outputs \cite{Oh2019}. A metamodel is trained on the observed set of outputs to predict the model architecture and the queries and attack strategies can be optimized using game theoretic solutions. 

The above attacks are mostly concerned with stealing the exact parameters and architecture of ML models via side channels and query optimization. Reverse Engineering parameters of classical ML models during inferencing on untrusted third-party cloud providers is generally not an issue as it is usually avoided by sandboxing the model and providing client access through higher-level APIs. However, such flexibility is not available in the quantum domain as the QML model is a quantum circuit that needs to be executed on the quantum hardware. Blind computation \cite{blind} is a viable direction for creating APIs for QML models, but it increases the overhead of communication severely by incorporating extra quantum states, operations, and qubits. Also, they are prone to decoherence errors and background noise of the quantum hardware. Although recent literature \cite{kbq} assumes that adversaries sharing the untrusted cloud providers with users can steal IP such as training data, we note that such attacks are not straightforward as even recovering rotation of a single-qubit rotation gate is non-trivial due to transpilation process.

\section{Threat Model and Analysis}

\subsection{Threat Model}
We assume that the quantum cloud vendor or a malicious entity within the vendor is untrustworthy or at the very least honest but curious. They may not alter the QML circuit or its outcome but may be interested in making a profit or just gaining deeper insight into the model of the victim. This is true since trained QML models are extremely expensive and valuable (as pointed out in Section \ref{intro}). For profit-making, an adversary would have to offer their own services using the stolen model. With access to the transpiled copy of the QML circuit during inference, the adversary can strip off the state preparation circuit and reuse the parametric part of the QNN. Note, that we assume that the QML model is trained on non-proprietary hardware which makes the design of the quantum circuit and training parameters valuable. To gain profit from the model, he will attach the transpiled version of the new state preparation circuit corresponding to the new inference data and execute it on the hardware that was used to transpile the original QML model. However, this will restrict the benefit of the model since it cannot be executed on other hardware. Furthermore, the adversary will need to know the logical to physical qubit mapping to attach the state preparation circuit correctly. Having access to the pre-transpiled version of the model is attractive from several perspectives. First, the adversary can transpile the model to any quantum hardware and qubit technology increasing the sell-ability of the stolen model. Second, the adversary can avoid legal issues by identifying, removing, or tampering with any possible embedded watermark in the original model or embedding their own watermark. Moreover, third, the adversary can refine the model for their target application by training it further, if needed. RE of the whole QML model is a multi-step process however, we focus on the recovery of the parametric rotation gates in this paper as the first step with the quantum classifiers as a test case. The objective of the attacker is to guess the original rotations such that the transpiled copy of the reverse-engineered circuit matches the original transpiled circuit closely.          



\subsection{Adversary Capabilities}
We assume the untrusted third-party provider possesses: 
(i) access to the white-box version of the transpiled version of the circuit of the QML model. This will act as the golden model that will be used to validate his guess about the rotation values of the parametrized gates, (ii) the transpiler which can be used to transpile the RE version of the model and validate their guess, (iii) substantial computational resources at their disposal to accelerate the search for parameters enabling quickly and reduce the error between original and RE'ed models, (iv) historical data, logs, and usage patterns of the QML model, which can be leveraged to gain additional insights into the hyperparameters of the QML model.


\section{Proposed Idea}
A PQC has two main components, the entanglement and the rotation gates. We attempt to extract the original architecture of the QML model of the user by identifying the entanglement and the type of rotation gates and follow it up by trying to determine the original parameters from the transpiled circuit. For the following setup, we consider the basis gate set of \texttt{[id, x, sx, cnot, rz]} that is common to almost all the IBM machines, which means that every logic gate of the circuit will be expressed as a combination of the gates from the basis gate set in the transpiled form of the user QML model.

\begin{figure*}
    
    \centering
    \includegraphics[width=1\linewidth]{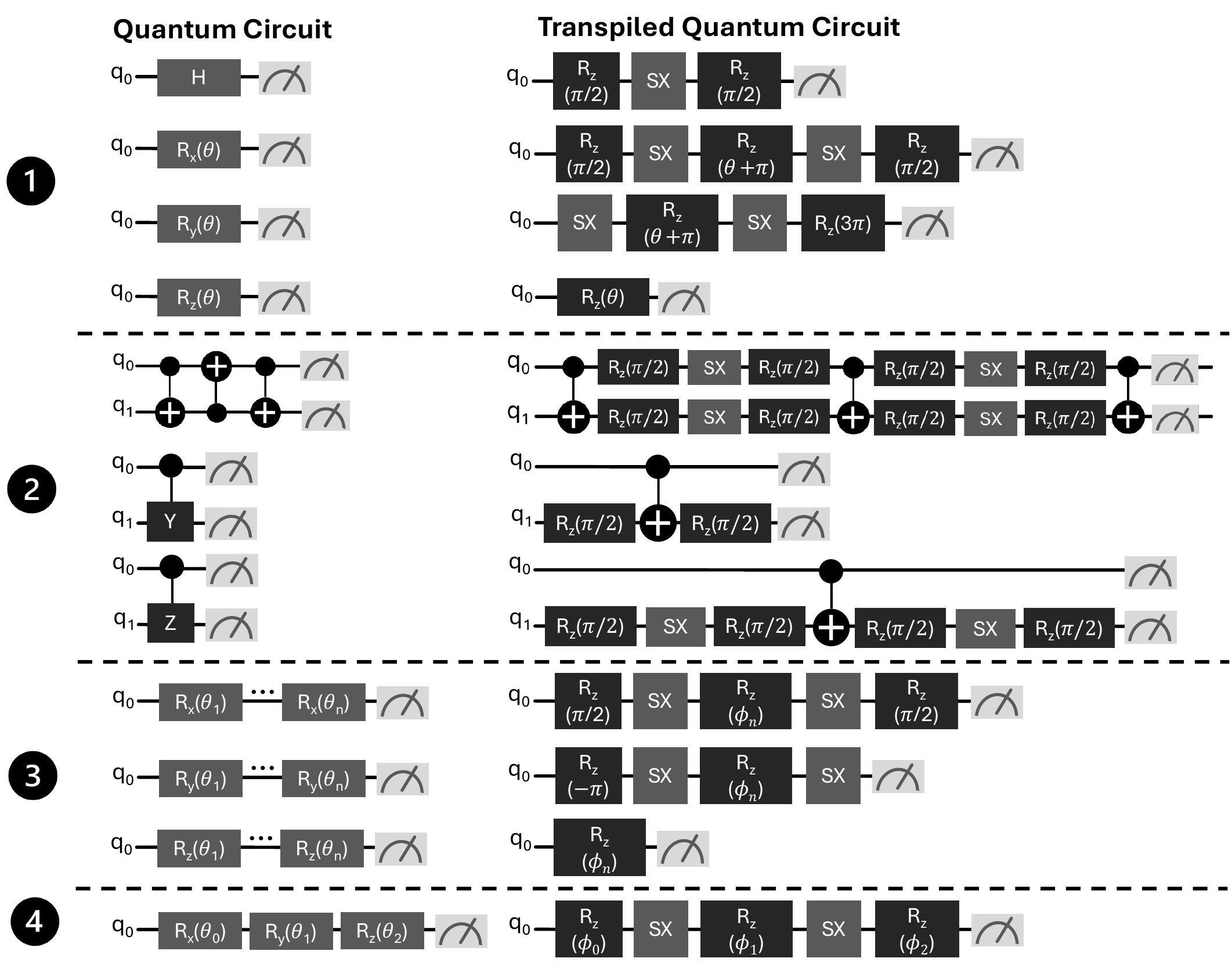}
    \caption{The adversary designs a Look-Up Table (LUT) based on basic circuit transpilations. The circuits shown here are transpiled on a backend having a linear coupling map with a basis set of \texttt{[id, x, sx, cnot, rz]} at an optimization level set to 1. In the diagram, (1) shows the transpilation of the Hadamard gate and the basic Rotation gates ($RX(\theta), RY(\theta), RZ(\theta) $); (2) shows the transpilation of basic 2-qubit entanglements. Since $CNOT$ is a part of the basis set, it remains as is and the other gates ($CY$, $CZ$) get transpiled into a combination of the basis gates; (3) shows the transpilation of a combination of multiple $RX(\theta)$, multiple $RY(\theta)$, and multiple $RZ(\theta)$ gates. They can be reversed into a single parameter of the corresponding rotation gate; and (4) shows the transpilation of a combination of $RX(\theta)$, $RY(\theta)$, and $RZ(\theta)$.  } 
    \label{fig:LUT}
    \vspace{-10pt}
\end{figure*}

\subsection{Reversing Entanglement}
Strong Entanglement in PQCs is realized primarily by introducing cascading layers of $CNOT$, $CY$, and $CZ$ gates. The configuration of these 2-qubit gates comprise one component of the architecture of the PQC. \textbf{CNOT} gates are straightforward to reverse as they are included in the basis gate set. However, this is true only when the coupling is linear or fully connected. In the case of a T-shaped coupling map where the $CNOT$ exists between qubits that are not connected physically, there is a $SWAP$ gate that is broken down as in Fig. \ref{fig:LUT}(2). To obtain the exact order of the CNOT, the transpiled circuit is parsed and the connections are identified. \textbf{CY} gates are transpiled as a combination of a $CNOT$ gate sandwiched between an $RZ(-\pi/2)$ gate and an $RZ(\pi/2)$ gate (Fig. \ref{fig:LUT}(2)). In a similar fashion, the \textbf{CZ} is transpiled as a $CNOT$ gate between two Hadamard ($H$) gates (Fig. \ref{fig:LUT}(2)).

The procedure to obtain the original architecture in terms of the arrangement of these gates involves parsing the transpiled circuit and obtaining the arrangement of the $RZ(\theta)$ and the $SX$ gates for every qubit and then using the LUT to identify the type of 2-qubit gate used (Algorithm \ref{algo:RE}).

\subsection{Identifying Original Parametric Gates}
The parameterized gates in a PQC provide finer-grained control over the quantum state by allowing the exploration of a larger portion of the Hilbert space using the rotation angles. Primarily, rotations in the $x$, $y$, and $z$ direction produce rotation gates $RX(\theta)$, $RY(\theta)$, and $RZ(\theta)$ respectively. Of these gates, we can find the $RZ(\theta)$ in the basis gate set of IBM machines. Therefore, the transpilation procedure of the rotation gates involves the expression of $RX$ and $RY$ as some combination of $RZ$ and some other gate from the basis gate set. 
\subsubsection{RX gates:} $RX$ gates can be represented as a combination of $RZ$ and Hadamard gates:
    
    \[
    \begin{split}
    RX(\theta) & = H \cdot RZ(\theta)  \cdot H \\ & = 
    RZ(\pi/2) \cdot SX\cdot RZ(\pi+\theta) \cdot SX \cdot RZ(\pi/2)
    \end{split}
    \]
The transpilation of the same can be observed from Fig \ref{fig:LUT}(1). We can identify potential $RX$ gates from the pattern of gates ($SX$ and $RZ$) in the transpiled circuit. It is observed that when the $RX$ gate is transpiled, the starting and the ending parameters of the $RZ$ gate are $\pi/2$. However, it is also observed that multiple $RX$ gates with the same or different parameters when stacked together produce the same order of gates with different parameter values after transpilation:
    \[
    \begin{split}
    RX(\theta_1)...RX(\theta_n) & = H \cdot RZ(\theta')  \cdot H \\ & = 
    RZ(\pi/2) \cdot SX\cdot RZ(\phi) \cdot SX \cdot RZ(\pi/2)
    \end{split}
    \]
Therefore, while reversing, we consider a single $RX$ gate since it is easier to obtain the parameter for a single $RX$ gate and it also reduces the number of parameters in the reversed circuit. 
\subsubsection{RY gates:}
$RY$ gates can be represented as a combination of $RX$ and PauliZ gates:
    \[
    \begin{split}
    RY(\theta) & = Z \cdot RX(\theta')  \cdot Z \\  
    \end{split}
    \]
After adjusting the global phase and expressing PauliZ and $RX$ gates as a combination from the basis gate set, $RY(\theta)$ can be represented as 
    \[
    \begin{split}
    RY(\theta) & = SX \cdot RZ(\theta+\pi) \cdot SX \cdot RZ(3\pi) \\  
    \end{split}
    \]

The transpilation is observed in Fig. \ref{fig:LUT}(1). However, we find the same pattern again for multiple $RY$ gates stacked together:
    \[
    \begin{split}
    RY(\theta_1)...RY(\theta_2) & = SX \cdot RZ(\theta') \cdot SX \cdot RZ(3\pi) \\  
    \end{split}
    \]
It can be observed that while transpiling the $RY$ gate, we obtain a pattern of $SX$ and $RZ$ gates which can be used to extract the gate and the parameter values from the transpiled circuit.

\subsubsection{RZ gates:} $RZ$ gates are a part of the basis set hence the presence of a single or a sequence of $RZ$ gates results in a single $RZ$ gate in the transpiled circuit.

\subsubsection{Multiple Rotation Gates:}
In a case where multiple rotation gates are stacked together, they get transpiled as a pattern $RZ(\theta_1) \cdot SX \cdot RZ(\theta_2) \cdot SX \cdot RZ(\theta_3)$, irrespective of the order and number of the $RX$, $RY$, and $RZ$ gates in the circuit. Therefore, on observing a similar pattern, we can reverse it to a combination of single occurrences of the three rotation gates:
    \[
    \begin{split}
    RZ(\theta_1) \cdot SX \cdot RZ(\theta_2) \cdot SX \cdot RZ(\theta_3) & = \\ 
    RX(\phi_1) \cdot RY(\phi_2) \cdot RZ(\phi_3)\\  
    \end{split}
    \]
Again, in this fashion, the number of parameters is reduced while reversing the circuit and obtaining a circuit that is architecturally closer to the QML model of the user with the new parameters functionally the same as the trained ones. 


\begin{algorithm}[t]
\caption{Reverse engineering of QML parameters}
\label{algo:RE}
\begin{algorithmic}[1]
\Procedure{parser}{$qc\_transpiled$, $LUT$}
    \State $param \gets [-\pi, \pi]$
    \State $temp \gets []$
    \For{$qubit$ in $qc\_transpiled$}
        \For{$gate$ in $qubit$}
            \State $temp \gets gate$
        \EndFor
        \State $qc\_new \gets \text{REVERSE}(temp, param, LUT)$
    \EndFor
    \State \textbf{return} $qc\_new$
\EndProcedure
\Procedure{reverse}{$gate\_list$, $param$, $LUT$}
    \State $gate \gets gate\_list.split[\text{CNOT}]$
    \State $temp2 \gets LUT(gate)$
    \State $temp2\_transpiled \gets temp2.transpile$
    \State \raggedright $diff \gets \Delta(temp2\_transpiled.param,$ 
    \linebreak $qc\_transpiled.param)$
    \For{$d$ in $diff$}
        \If{$d$ = $\min(diff)$}
            \State \textbf{return} $temp2, temp2\_transpiled.param$
        \EndIf
    \EndFor
\EndProcedure
\end{algorithmic}
\end{algorithm}

\subsection{Extracting Parameters}
While extracting the parameter values during the reverse engineering procedure, we note that the total phase of the circuit is always less than $2\pi$. Therefore, we start by reducing the search space for the parameters of the rotation gates to $[-\pi, \pi]$. Once, the transpiled circuit is parsed and the pattern of rotation gates is identified, a naive approach to determining the parameters is to perform a brute force search in $[-\pi, \pi]$ with a certain step size. The smaller the step size, the lesser the parameter estimate error. As observed in Algorithm \ref{algo:RE}, we define two functions, $parser$, and $reverse$. In the $parser$ function, we define $param$ as a list of all values between $-\pi$ and $\pi$ with a user-defined step size of $N$. We parse the transpiled circuit with respect to every qubit and isolate the associated gates. We pass this list of gates to the $reverse$ function that refers to the LUT to identify which corresponding combination of gates suits best for the qubit and transpile it for every possible combination from $param$ comparing the difference between the parameters of the original transpiled circuit and the reverse-engineered transpiled circuit. The closest set of parameters in the reverse-engineered transpiled circuit is the one where the difference between the parameters is minimal. The LUT is designed to have the transpilation of basic one-qubit and two-qubit gates but can be extended for more complex operations. Since the extraction of parameters is done by transpiling the circuit for every parameter till the closest set is obtained, the time complexity of the algorithm turns out to be $O(k^{2\pi/N})$; where $k$ is the number of parameters in the corresponding combination of gates from the LUT, and $N$ is the step size.

\section{Results}
\subsection{Simulation Setup}
\textbf{Traning: }We tested the idea of reverse engineering on multiple QML models to extract their parameters. The QML models have been implemented in Pennylane \cite{bergholm2022pennylaneautomaticdifferentiationhybrid} to utilize the \texttt{lightning.qubit} feature for performing linear algebra calculations faster. All QML models have been trained using the Gradient Descent Optimizer with a learning rate of 0.05, and a Mean Squared Error loss function has been used to evaluate the performance. The transpilation of the circuits for the QML models has been done using the transpiler library of Qiskit \cite{qiskit2024} keeping a linear coupling map, and a basis gate set of \texttt{[id, x, sx, cnot, rz]}. The reverse engineering of the transpiled circuits to extract the parameters has been done on the same setup as the transpilation procedure, running on a machine with 16GB RAM on an Intel Core i7-6700 CPU at a clock frequency of 3.40 GHz. 
\textbf{Dataset:} We conduct our experiments on the MNIST dataset \cite{mnist} picking labels as per the capacity of the QML model as a proof-of-concept of our approach. 

\subsection{Reversing QNNs}
We elaborate on a few examples to validate our concerns about the untrusted third-party vendor performing a reverse engineering operation of the transpiled circuits of the user to extract the parameters and a QML model that performs as well as the user-designed QML model. 
\begin{figure}
    
    \centering
    \includegraphics[width=0.8\linewidth]{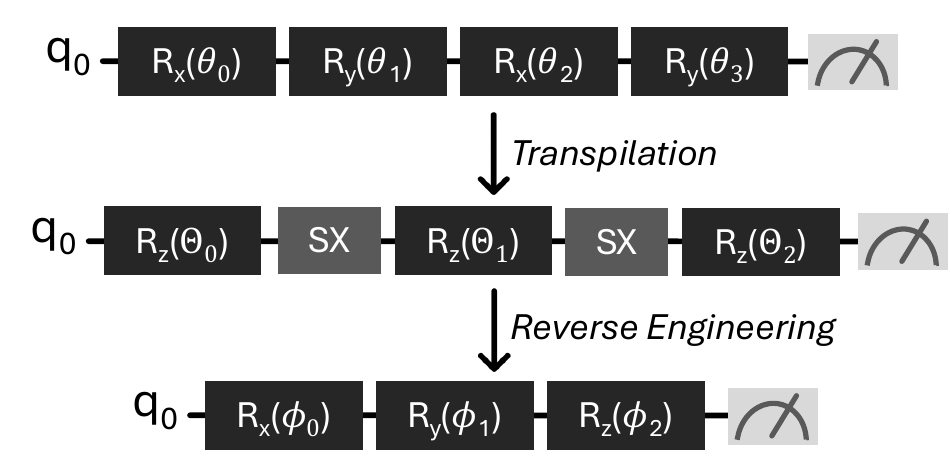}
    \caption{Diagram representing a 1-qubit classifier. It gets transpiled post-training and gets reverse-engineered to a classifier having three parameters. Both the user-designed and the reverse-engineered classifier show the same training accuracy. } 
    \label{fig:eg1}
    \vspace{-10pt}
\end{figure}
\begin{figure*}
    
    \centering
    \includegraphics[width=1\linewidth]{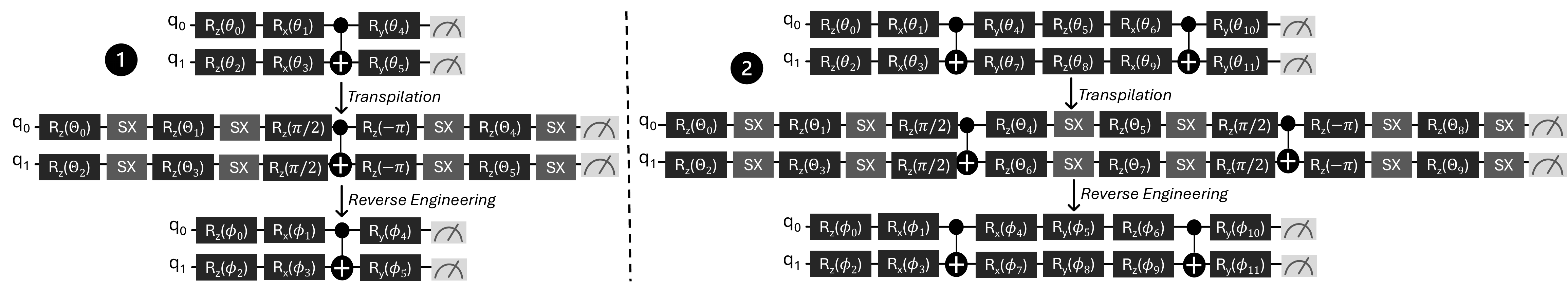}
    \caption{Transpilation and reverse engineering of the 2-qubit classifier. (1) represents a single layer of the classifier. The transpiled form gets reverse-engineered by the adversary using the LUT to obtain a circuit with a similar number of parameters as the original circuit. (2) shows two layers of the same circuit design. Here the adversary performs the reverse-engineering using a different case from the LUT to obtain a circuit with the same number of parameters. The reverse-engineered circuits show a minimal drop in training accuracy compared to the original model.} 
    \label{fig:eg2}
    \vspace{-10pt}
\end{figure*}
\textbf{Example 1: }We design a 1-qubit classifier to train it on the 0 and 1 labels of the MNIST dataset to perform binary classification on the data. To evaluate the efficacy of the reverse engineering procedure on the 1-qubit classifier, we obtain the transpiled circuit of the QML model as a QASM file. We can strip off the state embedding and obtain the transpiled circuit as observed in Fig. \ref{fig:eg1}. We parse this circuit using Algorithm \ref{algo:RE} and find the order of basis gates to be $RZ(\theta_1) \cdot SX \cdot RZ(\theta_2) \cdot SX \cdot RZ(\theta_3)$ which matches to (4) in Fig. \ref{fig:LUT}. Further, we apply the corresponding circuit to reverse engineer a set of parameters for the combination $RX(\phi_1) \cdot RY(\phi_2) \cdot RZ(\phi_3)$. On comparing the training details we can see, that the original circuit has a training accuracy of $\sim$70.29\%, and the training accuracy of the reverse-engineered QML model after transpilation is almost the same with an error of the order $10^{-16}$.

\textbf{Example 2: }In this scenario, we consider a 2-qubit classifier and investigate the reverse engineering in two cases-- when the circuit has (i) one layer, and (ii) two layers (Fig. \ref{fig:eg2}). We follow similar steps as in Example 1 to reverse when the circuit has one layer. We obtain the transpiled circuit of the QML model in the form of a QASM file, parse it, match the order of gates qubit by qubit using the $LUT$, and obtain the parameters. However, in the second case we find that on repeating a layer, the transpilation procedure combines the rotation gates between the $CNOT$ gates thus modifying the original architecture of the QML model designed by the user. To reverse engineer this circuit and extract the parameters, we take a similar route. However, this time we obtain a circuit that is not exactly the same based on design but has an equal number of parameters. Also, comparing the training accuracy of both we find that the original model has a training accuracy of $\sim$69.47\% and the reverse-engineered QML model has an accuracy of $\sim$67.22\% which is a loss of 3.2\% in accuracy and a mean error of $6.10 \times 10^{-2}$ in the reverse engineered parameters.

    

\begin{table}[h]
\caption{Error between the original and reverse-engineered classifiers ($i$-qubit, $j$-layer) }
\centering
\begin{tabular}{c||c c c c}

    \multirow{2}{*}{\textbf{Classifier}} & \multirow{2}{*}{\textbf{\#Params}} & \multicolumn{2}{c}{\textbf{Parameter}}  & \multirow{2}{*}{\textbf{Acc.Error \%}}\\
    & & \textbf{Mean} & \textbf{SD} \\
    \hline
    \hline
    1Q & 4  & 5.94e-02 & 8.55e-02 & 1e-16\\
    \hline
    2Q; 1-layer & 6 & 5.33e-02 & 2.50e-02 & 1.7\\
    2Q; 2-layer & 12 & 6.10e-02 & 4.43e-02 & 3.2\\
    2Q; 3-layer & 18 & 8.45e-02 & 8.99e-02 &  5.7\\
    \hline
    4Q; 1-layer & 8 & 7.29e-02 & 7.73e-02 & 2.1\\
    4Q; 2-layer & 16 & 9.29e-02 & 9.91e-02 & 5.9\\
    4Q; 3-layer & 24 & 1.18e-01 & 9.79e-02 & 6.3\\
    \hline
    8Q; 1-layer & 16 & 6.16e-02 & 3.84e-02 & 4.1\\
    8Q; 2-layer & 32 & 8.71e-02 & 3.69e-02 & 5.3\\
    8Q; 3-layer & 48 & 1.71e-01 & 2.81e-01 & 7.6\\
    \hline
    \hline
\end{tabular}

\label{table:error}
\end{table}

\subsection{Error Analysis}
We measure the efficacy of the reverse engineering procedure by calculating the difference in the parameters of the transpiled circuit of the user-designed QML model and the reverse-engineered circuit of the QML model. We report this as the error value for the reverse-engineered circuit. A lesser error indicates that the reverse-engineered circuit is architecturally and functionally closer to the original circuit. Table \ref{table:error} shows the mean and standard deviation of the error values of the parameters. We also calculate an error as the percentage decrease in the testing accuracy of the classifier that is observed between the user-designed QML model and the reverse-engineered QML model. We can observe a considerable increase in the mean error while extracting the parameters by reverse engineering as the circuit design involves more qubits while maintaining a decently close training accuracy with the user-deigned QML model. This increase is justified in the sense that the error incurred while reverse engineering one parameter gets accumulated over the entire design of the circuit, thus increasing with the increase in complexity and number of parameters.

\begin{figure}
    
    \centering
    \includegraphics[width=0.8\linewidth]{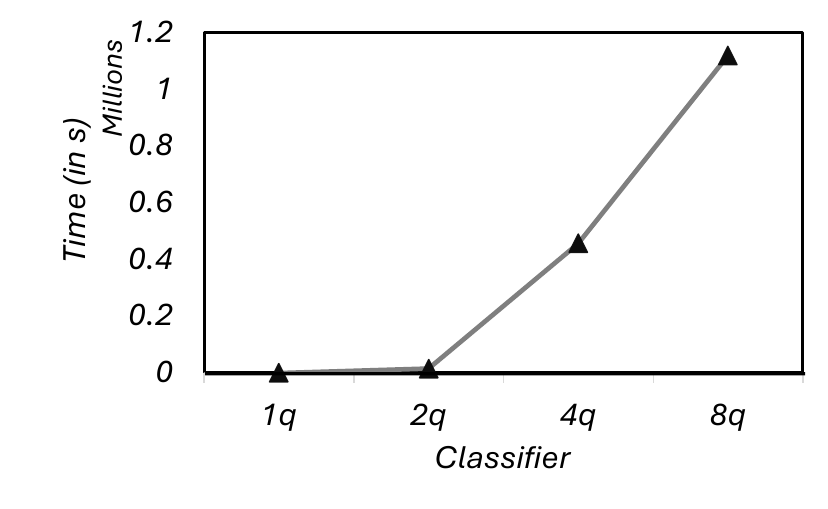}
    \caption{Plot demonstrating the time taken to reverse engineer QML classifiers. The X-axis represents the number of qubits in the classifier design. These classifiers have 3 layers. It can be observed that the overhead for reverse engineering QML classifiers is significantly high for a higher number of qubits.} 
    \label{fig:plot1}
    \vspace{-10pt}
\end{figure}

\begin{table}
\caption{Effect of step size on error and reverse engineer time for a 1-qubit classifier}
\centering
\begin{tabular}{c||c||c}
\textbf{Step size} & \textbf{Time (in s)} & \textbf{Mean error} \\ \hline \hline
1e-01 & 51 & 5.94e-02  \\ 
1e-02 &  6.62e+02 & 3.66e-03 \\ 
1e-03 & 8.52e+05 & 5.01e-04\\  
\hline \hline
\end{tabular}
\label{tab:step}
\end{table}

\subsection{Overhead Analysis}
We analyze the overhead incurred by the adversary while extracting the parameters using reverse engineering. 
From Fig. \ref{fig:plot1}, we can observe that although the time taken to reverse engineer a circuit representing QML classifiers increases considerably with the increase in the number of qubits, it is quite possible to perform successful reverse engineering to a certain degree keeping the mean error of the extracted parameters low. We also analyze the time taken to extract the parameters by reverse engineering the classifier using a lower step size in the brute force approach thus increasing the granularity of the search. We perform a set of experiments on the 1 qubit classifier (Fig. \ref{fig:eg1}) and present the results in Table \ref{tab:step}. We note that the time increases exponentially with reduced step size even by a small amount (=0.1) and the corresponding decrease in the mean error of the parameters observed post the reverse engineering procedure is not comparable. Therefore, considering the low error and difference in the accuracy of testing of the original and the reverse-engineered circuit of the QML model, we conclude that even a larger step size of 0.1 is sufficient to extract the parameters.

\begin{table}
\caption{Increase in time taken to RE a 4-qubit classifier on increasing the layers}
\centering
\begin{tabular}{c||c}
\textbf{\# Layers} & \textbf{Time (in s)} \\ \hline \hline
1  & 1.51e+04  \\ 
2 & 2.81e+04  \\ 
4 & 7.75e+05 \\  
8 & $>$ 1e+06\\
16 & $>$ 1e+07 \\ \hline \hline
\end{tabular}
\label{tab:layer}
    \vspace{-10pt}
\end{table}

\subsection{Considerations for Noise}
The experiments have been performed on noiseless simulators. Inherent noise in quantum hardware directly affects the parameterized rotation gates during the training phase of a QML model. However, in the attack model, the adversary obtains the trained model and performs RE on it to extract the trained parameters. Therefore, the RE procedure is unaffected by noise in quantum hardware and hence, the usage of noiseless simulations does not alter the concept of RE and the overhead analysis.

\section{Countermeasures}
We develop countermeasures against potential RE attacks on QML models based on two main observations: (i) The time taken to perform RE on a QML classifier increases with the number of layers in the circuit (Table \ref{tab:layer}), and (ii) The time taken to RE and the mean error in reverse-engineered parameters and the testing accuracy increases with the number of parameters in the QML circuit (Table \ref{table:error}). We perform experiments to develop countermeasures considering a baseline 2-qubit QML classifier with three layers.

\subsection{Increasing the number of layers}
\begin{figure}
    
    \centering
    \includegraphics[width=\linewidth]{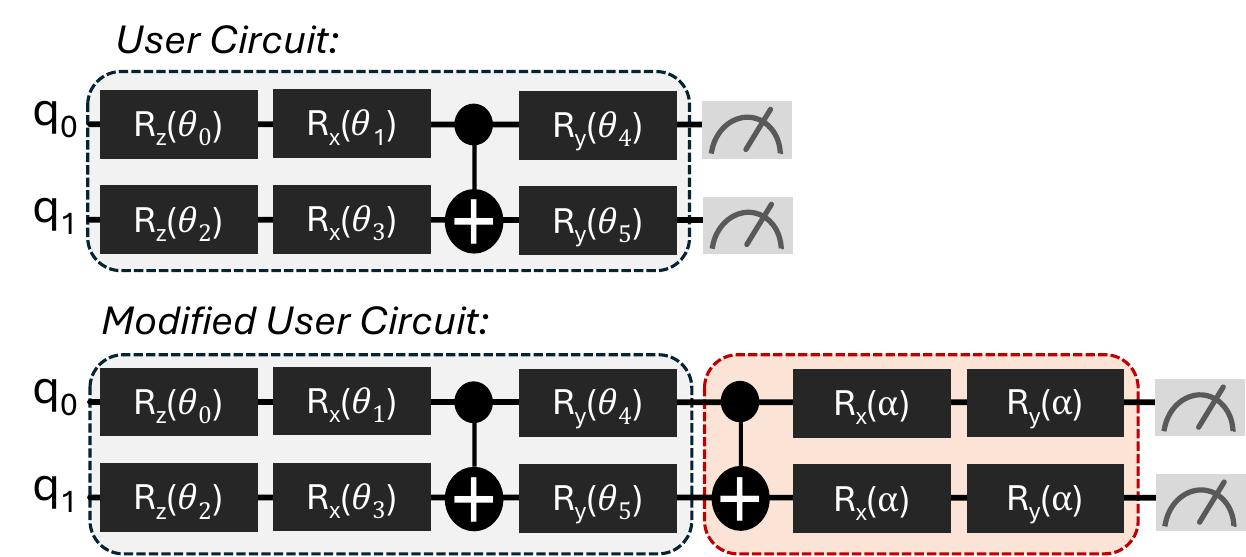}
    \caption{Adding an extra layer of rotation gates with fixed parameters to the existing QML model to resist RE attack. The shaded portion in red represents the fixed parameters which get transpiled as rotation gates forcing the adversary to RE them. This approach increases the overall time for the parameter extraction.}
    \label{fig:cm1}
    \vspace{-10pt}
\end{figure}

We propose to add dummy rotation gates with fixed parameters to the user-designed QML model, alongside the trainable parameters. This approach aims to increase the RE effort without hurting training time significantly. On transpiling the modified circuit, the trainable and fixed parameters get optimized together making it impossible for an adversary to distinguish between them. The adversary would try to reverse engineer all the parameters considering them to be trainable, thus increasing the overhead of the RE significantly. In Fig. \ref{fig:cm1} example, the user circuit has six trainable parameters and the modified circuit has ten (six trainable, four fixed). From Fig. \ref{fig:cmplot}(1), we can see the drastic increase in RE time with an increase in the number of layers in the classifier. The user can choose to repeat the layers of trainable parameters to increase the granularity of the QML model and increase the layers of the fixed parameters to make the model RE resistant. 

\subsection{Increasing the number of qubits}
\begin{figure}
    
    \centering
    \includegraphics[width=0.7\linewidth]{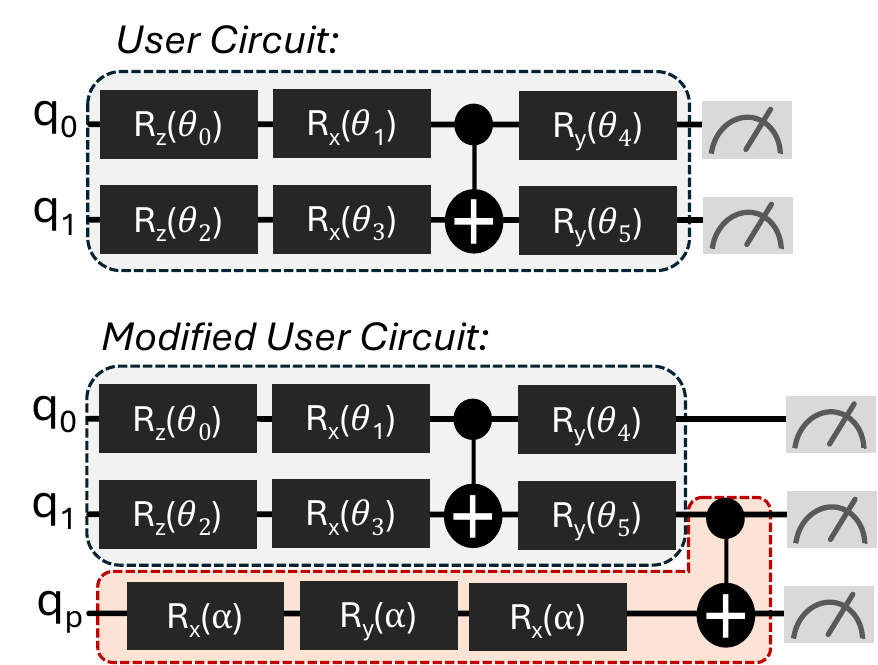}
    \caption{Adding extra qubits with fixed parameters to resist RE attack. The three rotation gates with fixed parameters are added to the existing QML model to increase the RE time.}
    \label{fig:cm2}
    \vspace{-10pt}
\end{figure}

Another approach to resist RE without affecting training time significantly is by adding dummy qubits with fixed parameters. The rotation gates with fixed parameters get transpiled as normal rotation gates making it indistinguishable from the trainable parameters. In Fig. \ref{fig:cm2} example, the user adds an extra qubit, $q_p$ to the 2-qubit classifier with three fixed parameters. From Fig. \ref{fig:cmplot}(2), we obtain a $3.41 \times$ increase in the overhead on adding 8 qubits with fixed parameters to the classifier. The user can increase the number of layers with the trainable parameters to make the classifier better and simultaneously increase the number of qubits with fixed parameters to resist RE attacks.

\begin{figure}
    
    \centering
    \includegraphics[width=\linewidth]{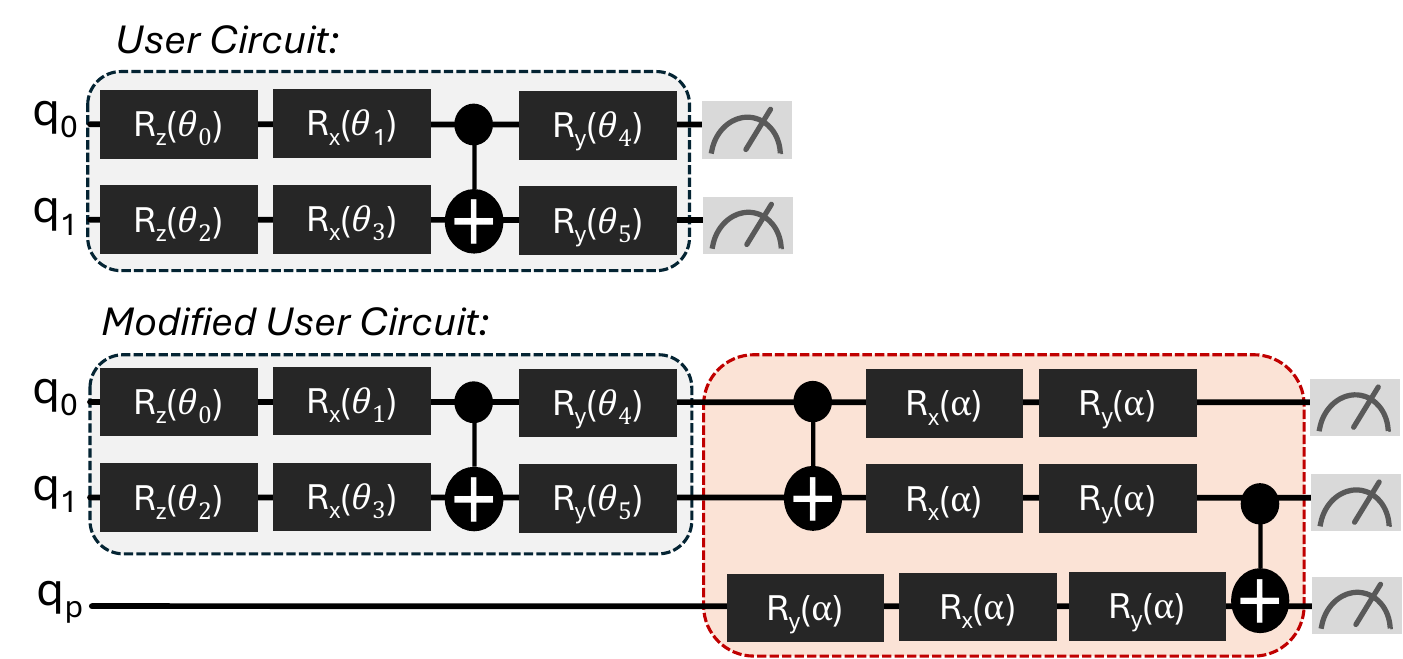}
    \caption{Addition of an extra qubit $q_p$ as well as an extra layer of fixed parameters (shaded in red) for RE resistance. The adversary needs to RE and extract 13 parameters instead of the 6 trainable ones which increases the overhead significantly.}
    \label{fig:cm3}
    \vspace{-10pt}
\end{figure}

\subsection{Combining dummy layers and qubits}
\begin{figure}
    
    \centering
    \includegraphics[width=0.9\linewidth]{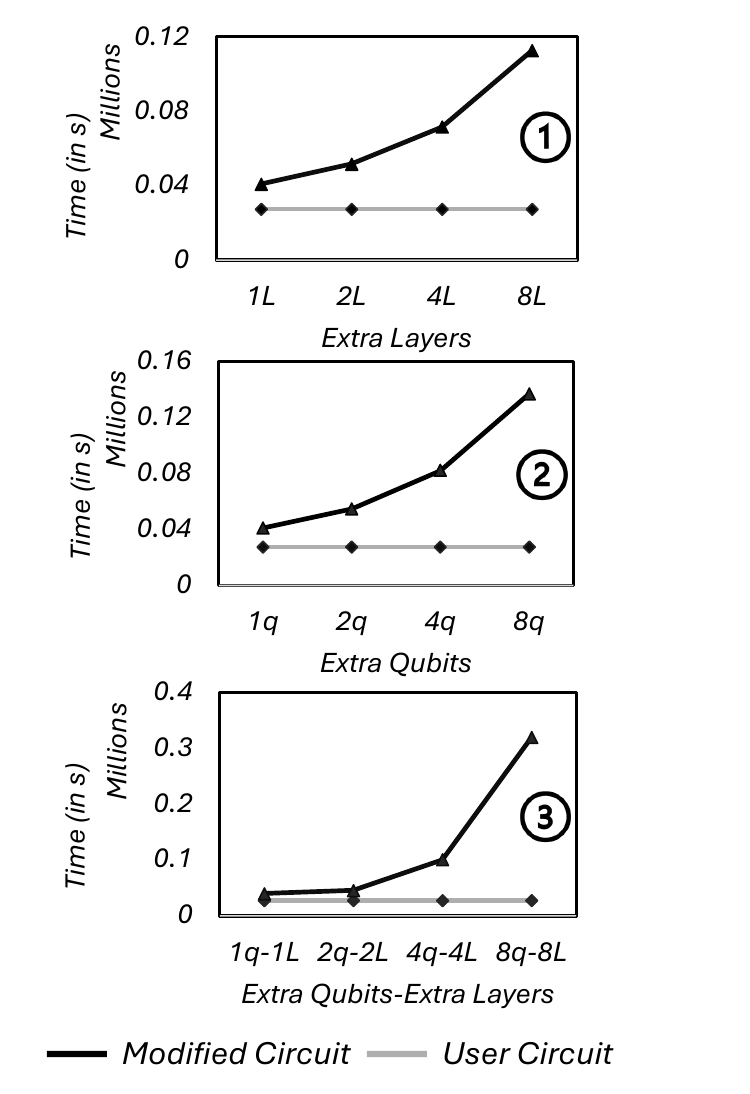}
    \caption{Plots representing the difference in time taken to RE the QML circuit before and after adding the fixed parameters. The experiments are done on an existing user QML model of 2 qubits and 3 layers. We can observe from the plots that adding fixed parameters to extra qubits and layers together will increase the overhead more, thus providing better security. }
    \label{fig:cmplot}
    \vspace{-10pt}
\end{figure}

To obtain a higher level of security against RE, the user may opt to add dummy qubits as well as dummy layers to his classifier which will increase the number of fixed parameters. This way the user can keep the circuit design compact by making a conscious balance between the number of qubits and layers instead of adding only extra qubits or layers. In Fig. \ref{fig:cm3} example, the user adds a layer of fixed parameters alongside an extra qubit, $q_p$ to the existing 2-qubit classifier. The adversary in this case has to RE 13 parameters instead of the 6 trainable ones. From Fig. \ref{fig:cmplot}(3), we observe an almost exponential growth in RE time. Similar to the above-mentioned approaches, the user can choose to increase the layers of fixed parameters and the number of dummy qubits without affecting the training overhead.

\subsection{Overhead analysis}
We analyze the impact of the proposed countermeasures on the training performance of the original QML model. For a case study, we consider a baseline QML model with two qubits and three layers (Fig. \ref{fig:cm1}) that has 18 trainable parameters. We consider all three cases, adding a layer of 4 fixed parameters (Fig. \ref{fig:cm1}), adding a dummy qubit with 3 fixed parameters (Fig. \ref{fig:cm2}), and adding a dummy qubit and an extra layer with seven parameters (Fig. \ref{fig:cm3}) and compare their performance with the user circuit. As observed in Table \ref{tab:perfov}, there is a slight decrease in the performance of the modified circuits on adding a dummy qubit or an extra layer.

\begin{table}
\caption{Analysis of the performance overhead on a 2-Q, 3-layer classifier on including the countermeasures}
\centering
\begin{tabular}{c||c}
\textbf{Modification type} & \textbf{\%Difference in Acc.} \\ \hline \hline
Dummy Qubit  & 8.76  \\ 
Extra Layer & 7.01  \\ 
Extra Layer and Dummy Qubit & 3.53 \\  
 \hline \hline
\end{tabular}
\label{tab:perfov}
    \vspace{-10pt}
\end{table}

\section{Conclusion}
We explore reverse engineering (RE) of transpiled Quantum Machine Learning (QML) circuits as an attack model by untrusted third-party cloud providers. We propose an approach to perform RE on QML circuits and extract the parameters. We test the efficacy of our approach by training the models and performing RE on them and conclude from the results that reverse-engineered QML models can achieve training accuracies nearly identical to the original models in a reasonable time, underscoring the severity of the threat. We also include countermeasures like adding fixed, non-trainable parameters to the QML circuit design that increase the overhead of RE significantly for the adversary, which users can adapt to protect the IP of their design.

\begin{acks}
The work is supported in parts by the National Science Foundation (NSF) (CNS-1722557, CCF-1718474, DGE-1723687, and DGE-1821766) and gifts from Intel.
\end{acks}

\bibliographystyle{unsrt}
\bibliography{refs}


\end{document}